# Comparative Mathematical Study of Blood Flow Through Stenotic and Aneurysmatic Artery with the Presence and Absence of Blood clots.

**Mohammed Nasir Uddin[*1] M. Monir Uddin[2] Md. Monjarul Alam[3]**

[1]Department of Information and Communication Technology (ICT), Bangladesh University of Professionals (BUP), Dhaka-1216, Bangladesh
Email: nasirbuet@gmail.com

[2]Department of Mathematics and Physics, North South University, Dhaka-1229, Bangladesh
Email: monir.uddin@northsouth.edu

[3]Dept. of Electrical and Electronic Engineering, Islamic University, Kushtia-7003, Bangladesh
Email: mdmonjarul.alam2@mail.dcu.ie

*Abstract:*

Numerical predictions of blood flow and hemodynamic properties through a stenotic and aneurysmal rigid artery are studied in the presence of blood clot at constricted area. Finite element method has been used to solve the steady partial differential equations of continuity, momentum, Oldroyd-B and bioheat transport in two dimensional cartesian coordinates system.The present investigation carries the potential to compute blood velocity, pressure and drag coefficient with major significance at the throat of stenosis and aneurysm. The models are also employed to study of simulation, influence of blood clot and hemodynamical characteristics for all modifications. The back flow and recirculation zones are found at stenotic and aneurysmal region for the model. The quantitative analysis is completed by numerical calculation having physiological significance of hemodynamical factors of blood flow depends on the dimensionless parameters which show the validity of present model.

## 1. Introduction

Cardiovascular system of human body is affected due to irregular cell increases or decreases in blood vessel which leads to arterial diseases in various organs. Although it is not clearly identified of exact mechanisms for invent of this matter, but is found that blood flow disorder, initiation of arterial diseases and irregularities have been created from a mild stenosis [1,2]. A good number of experimental, theoretical and numerical study has been continued for understanding of stenotic blood flow. Smith [3] has been studied details of steady flow through an axisymmetric stenosis and show that blood flow strongly depends on the stenotic geometry condition. The real understanding that the blood pulsatile flow cannot be ignored, many laboratory work, theoretical study and computation analyses on the blood flow have been completed at stenotic conditions for Newtonian fluid [4-8]. It is very important that the blood flow will obey Newtonian law for high shear rate flow. In case of low shear rate, the fluid behavior will be non-Newtonian specially if the blood flow is in smaller arteries and in the down steam of the stenosis. It has been pointed out that blood shows non-Newtonian performances (Thixotropy, Viscoelasticity and Shear thinning) for cardiovascular diseases. Recently researchers [9] has been given more concentration for non-Newtonian blood flow analysis with various stenosis conditions and still have a lot of scope to work in this area.

In this paper, we also consider aneurysm condition of blood vessel, it is another severe situation that about 75 percent of all patients have died before reaching the hospital and it is extreme risk for human body [10]. Ingoldby et al. [11] have shown these prehospital deaths are counted, the overall mortality rate may exceed 90 percent. Ernst [12] has studied that the increasing median age of the population contributes to an increasing incidence. It is important to identifying the risk features that may have key role in aneurysm growth and rupture has become a unified multidisciplinary task to clear understanding on the pathogenesis and growth of aneurysm. Numerical and experimental study has performed to interpret of blood flow through anuerysmatic artery during last and present decades [13-15] and biomechanics has been receiving the attention of researchers on this area and providing numerical solution of human blood flow system in various conditions [16]. In most of the inquiries, the flow is considered in cylindrical pipes with uniform cross-section area. But, is well known that blood vessels deformed at the change of bone shape for various reasons in human body. Hence the idea of blood flow in a various cross-section forms the prime basis of a large class of problems in understanding blood flow pattern. Most of the study have considered long, narrow and tapering vessels for blood flow analysis [17-19] of Newtonian and non-Newtonian case. Thus, the effects of stenotic and aneurysmatic vessel with the non-Newtonian performance of the flowing blood seem to be identical significant and hence definitely deserve special consideration.

Therefore, we are motivated for mathematical study of blood flow through stenotic and aneurysmatic artery with blood clot. Thus, an effort is made in the present theoretical study to develop a mathematical model and numerical analysis in order to investigate the significant characteristics of the blood flow through a rigid stenotic and aneurysmatic vessel. The generalized cross model is used for non-Newtonian behavior of flowing blood flow. Malek et al. [20] widely studied the existence and uniqueness as well as the stability characteristics of streaming flow problems. The present problem is one of the major physiological significance, due attention is also paid to comparable study in the presence of blood clot and without blood clot for considered model.

We have used the following notations which are available in mathematical structures. The dimensionless number Weissenberg number, Reynold number and Prandtl number are denoted by *Wi, Re* and *Pr* respectively, $\sigma$ denotes stress tensor, *P* is the pressure, and stress tensor is denoted by $\sigma$. The velocity components are *U* and *V* along X and Y respectively. The perfusion coefficient *is* $\rho_f$, $\mu_v$ is viscoelasticity components and heat source is denoted by *Q*.

## 2. Preliminaries

This section briefly reviews some fundamental concepts from the previous literatures. The goal is to establish some notations and results that will be used in the later section of this paper.

From the invent of stenosis or aneurysm effects on blood flow in our human arteries have created serious disorders in circulatory system that leads various cardiovascular diseases such as arteriosclerosis, bleeding, stroke, kidney damage etc. [21,22]. At present time medical researchers, bioengineers and numerical scientists join efforts with the purpose of providing numerical simulations of human blood flow system in different conditions. The exchange of knowledge and data information is main purpose of this kind of collaboration that can be used in simulations. A relation has been established by Thurston [23] between blood shear rate and viscoelasticity of blood flow. At first, Wille [24] has described numerical study of pulsatile blood flow through aneurysms vessel and shown existence of vortex with various size during the cardiac cycle. Oka [25] has studied a taper analysis of blood flow to identify arterial disease and shown the pressure development is a vital factor. Kumar et al. [26] have explained the pulsatile suspension flow in a dilated blood vessel. They have determined that the additional complicate of hemodynamics in diseased areterial vessels having aneurysm and stenosis artery. During blood circulation system, the hemodynamical features of blood can plays a vital role to develop sever fatal cardiovascular diseases. Anand et at. [27] has described a model to blood flow simulation with Oldroyd-B fluid properties. They have [28] also exposed a model with blood clots for the formation and analysis. Shih et al. [29] have studied on skin surface with sinusoidal heat flux condition. A thermodynamic framework has studied by Rajagopal and Srinivasa [21] to show the blood viscoelastic response with various configurations. With the presence of hematocrit in a local aneurysm, the systematic analysis of flow features through a tube and artery have studied by Mukhopadhyay and Layek [10]. Prokop and Kozel [30] have shown a numerical simulation of generalized Newtonian and Oldroyd-B fluid with an extended computation domain. Nadeem et al. [2] have shown the reduce of the wall shear stress and resistance impedance to blood flow with the influence of metallic nanoparticles. The pulsatile blood flow has explicated by Achab et al. [31] to evaluate the blood flow characteristics and the wall shear stress under physiological conditions through a stenosis artery. Recently, Nasir and Alim [32] have described the blood flow through stenotic artery with various flow rates numerically. Later they [33] have also demonstrated the numerical investigation of blood flow through double stenotic artery.

A large scale numerical analysis is carried out by performing mathematical computations of the desired quantities having more physiological significance to explore the effects of stenotic vessel, aneurysmatic artery, the blood clot, the drag coefficient, dimensionless numbers and the non-Newtonian behaviour of the streaming blood on the physiological flow phenomena which are extensively quantified through their graphical representations presented at result section with appropriate scientific discussions. A comparative study is exposed in the presence of blood clot and without blood clot to know the significances of coagulation of blood. A code validation results are also shown with Prokop et al [30] to authenticate the applicability of the present study.

## 3. Modeling and analysis of blood flow

This section contains main results of this paper. At first, we specify the physical model. Then the formulate the mathematical model using partial differential equations (PDE) and its boundary conditions. Finally, numerical technique is discussed to solve the mathematical model efficiently.

### 3.1 Model Specifications

The treated model is a two-dimensional rigid artery with stenosed and aneurysm vessel wall in cartesian system. The considered model in the present study is displayed in figure 1a. The figures 1b and 1c are main motivation for considering present model. The model is assumed with stenosed and aneurysmal hight $h_s$ and $h_a$ respectively where $h_s$=R and $h_a$=3R. The upper and bottom stenosed vessel walls are cool ($T_c$) and heated ($T_h$) while the rest of the walls are adiabatic and impermeable with no slip conditions. The velocity profile is prescribed at entrance and pressure is fixed to constant at outlet. The blood flow acceleration is expected at stenotic and aneurysmatic cross-sectional area. A mathematical structure has used by Achab et al [31] in cosine shaped for stenosis model. The modified mathematical model of physical construction in the presence of stenosis and aneurysm is in cosine shaped as follows

$$y = \begin{cases} 0.5D\left[1 - \frac{\varepsilon}{D}(1 + cos\pi\, M(x))\right]; 0 \leq x \leq L_s \\ 0.5D\left[1 + \frac{\varepsilon}{D}(1 + cos\pi\, N(x))\right]; L_s < x \leq L_a \\ 1 \qquad\qquad\qquad\qquad\qquad ; Other\ wise \end{cases}$$

Where $M(x) = x - \frac{x_s}{L}$ and $N(x) = x - \frac{x_a}{L}$, $x_s$ and $x_a$ are the center of stenosis and aneurysm respectively, L is the length of stenosis and aneurysm, $\varepsilon$ is the maximum height of stenosis and aneurysm, The model diameter, D=2R, The location of stenosis, $L_s = 3R$, the position of aneurysm $L_a$= 3R, total model length is 10R and R=3.1.

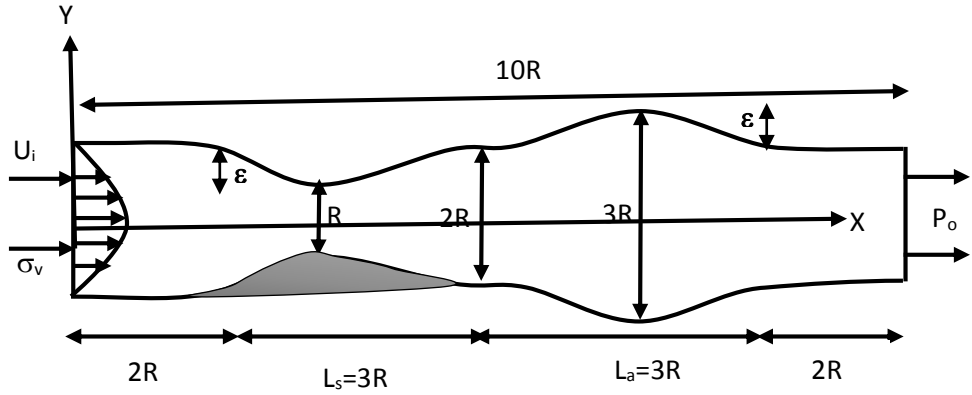

Fig.1a Structure of the computational domain

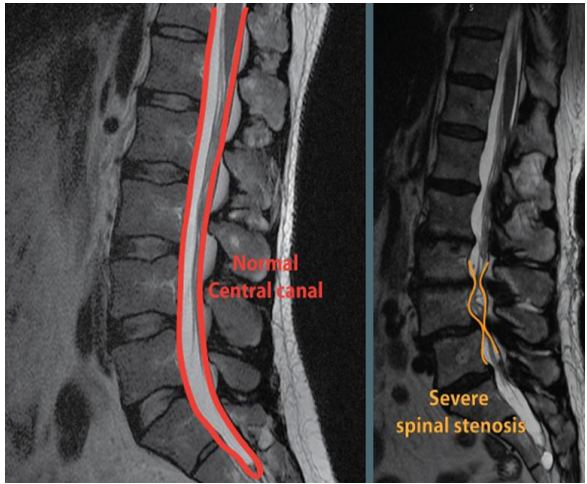 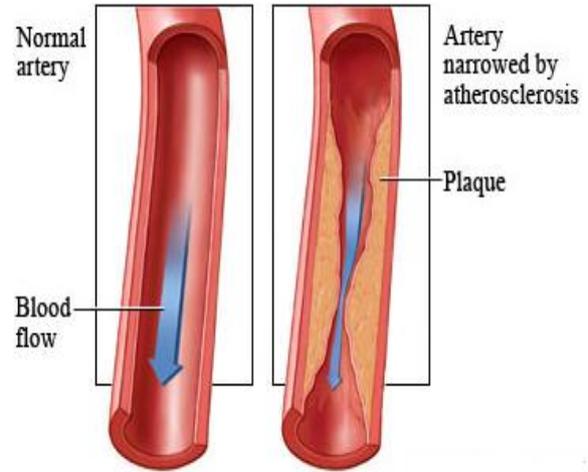

Fig.1b view of severe spinal stenosis [34]    Fig. 1c view of development of plaque [35]

## 3.2 Mathematical Model

The functioning fluid is assumed to be laminar blood flow and incompressible with shear-thinning and viscoelasticity properties. The generalized Oldroyd-B model is used to capture non-Newtonian properties of the blood flow. Using non-dimensional scale variables, the partial differential equations (1), (2), (3) and (4) are obtained in vector form for 2D stenosis and aneurysmal vessel in domain $\Omega$ as follows:

Continuity Equation:
$$\nabla \cdot U = 0 \tag{1}$$

Momentum Equations:
$$Re[(U.\nabla)U] = -\nabla P + (1-\lambda)\Delta U + \nabla.\sigma + f \tag{2}$$
Oldroyd-B constitutive equation:
$$W_i[(U.\nabla)\sigma] + \sigma = 2\mu_v V(U) + W_i[(\nabla U)\sigma + \sigma(\nabla U)^t] \tag{3}$$
Bio-heat Equation:
$$Re\,Pr(U.\nabla)\theta = \nabla^2\theta + Q - \rho_f\theta \tag{4}$$

Where, $U$ and $V$ are the velocity components along the $X$ and $Y$ axes respectively, $P$ is the pressure, $\sigma$ is stress tensor, $\rho_f$ is perfusion coefficient, and Weissenberg number ($Wi$), Reynold number ($Re$), Prandtl number ($Pr$) are dimensionless numbers. To non-dimensionalized the above equations we incorporating the dimensionless variables given below:

$x=LX$, $y=LY$, $t = Lt^*/U$, $u = UU_o$, $v = VU_o$ $p = \mu UP/L$, $\sigma = U\mu\sigma^*/L$, $f = f^*\mu U/L^2$, $\nabla = \nabla^*/L$, $Wi = \lambda_x U/L$, $Re = \rho UL/\mu$, $Pr = \mu c_p/k$, $\theta = k(T-T_c)/q_0L^2$, $q = Qq_0$, $w_b = \rho_f k / \rho c_b L^2$.

The boundary conditions are required to solve the governing equations and there are four disjoint parts of the boundary wall: rigid wall, outlet, blood clot and inlet. The axial velocity profile **u** is considered along x-axis and set zero velocity along y-axis at inlet artery. The extra stress components **T_e** and pressure **P_i** are prescribed at the inlet. A constant pressure **P_o** is imposed on the pressure at the outlet. On rigid walls, all velocity components are set to zero. On blood clot, the wall is heated and cooled at stenotic artery.

### 3.3 Numerical Solutions

The discretized momentum and Oldroyd-B equations subjected to the boundary conditions simultaneously will be solved using the Matlab programming [36] & Mathematical programming package COMSOLMULTIPHYSICS [37] or the dependent variables (velocity, pressure and stress tensor). The numerical procedure [38,39] have been used in this work is based on the Galerkin weighted residual method of finite element formulation and details are given here below. After the appling the weighted residual method then

$$\int_A N_\alpha(\nabla.U)dA = 0 \tag{5.1}$$

$$Re\int_A N_\alpha U(\nabla.U)dA = -\int_A H_\lambda \nabla P\,dA + \int_A N_\alpha \nabla\sigma\,dA$$
$$+ (1-\lambda)\int_A N_\alpha \nabla^2 U\,dA + \int_A N_\alpha f\,dA \tag{5.2}$$

$$Re\int_A N_\alpha U(\nabla.V)dA = -\int_A H_\lambda \nabla P\,dA + \int_A N_\alpha \nabla\sigma\,dA$$
$$+ (1-\lambda)\int_A N_\alpha \nabla^2 V\,dA + \int_A N_\alpha f\,dA \tag{5.3}$$

$$W_i\int_A N_\alpha U(\nabla.\sigma)dA + \int_A N_\alpha \sigma\,dA = \int_A \mu_v N_\alpha(\nabla.U)dA +$$
$$W_i\int_A N_\alpha(\nabla.U)\sigma + \sigma(\nabla.U)\,dA \tag{5.4}$$

$$Re\,Pr\int_A N_\alpha (U.\nabla)\,\theta\,dA = \int_A N_\alpha \nabla^2 \theta\,dA + \int_A N_\alpha Q\,dA$$
$$-\int_A N_\alpha \rho_f \theta\,dA \tag{5.5}$$

To generate the boundary integral terms associated with the surface tractions, extra stress tensor and temperature the Equations (5.2-5.5) become after appling Gauss's theorem.

$$Re\int_A N_\alpha U(\nabla.U)dA + \int_A H_\lambda \nabla P\,dA - \int_A N_\alpha \nabla.\sigma\,dA$$
$$-(1-\lambda)\int_A N_\alpha \nabla^2 U\,dA - \int_A N_\alpha f_x\,dA = \int_{S_0} N_\alpha S_x\,ds_0 \tag{5.6}$$

$$Re\int_A N_\alpha U(\nabla.V)dA + \int_A H_\lambda \nabla P\,dA - \int_A N_\alpha \nabla\sigma\,dA$$
$$-(1-\lambda)\int_A N_\alpha \nabla^2 V\,dA - \int_A N_\alpha f_y\,dA = \int_{S_0} N_\alpha S_y\,ds_0 \tag{5.7}$$

$$W_i \int_A N_\alpha U.(\nabla\sigma)\,dA + \int_A N_\alpha \sigma\,dA - \int_A \mu_v N_\alpha \nabla.U\,dA -$$
$$W_i \int_A N_\alpha [(\nabla.U)\sigma + \sigma(\nabla.U)]dA = \int_A N_\alpha \sigma_w\,ds_w \tag{5.8}$$

$$Re\,Pr\int_A N_\alpha (U.\nabla)\theta)\,dA - \int_A N_\alpha \nabla^2 \theta\,dA - \int_A N_\alpha Q\,dA$$
$$+\int_A N_\alpha \rho_f \theta\,dA = \int_{s_w} N_\alpha q_w\,ds_w \tag{5.9}$$

Here Equations (5.2-5.3) specify surface tractions ($S_x$, $S_y$) along outflow boundary $S_0$, Equations (5.4-5.5) specify velocity components, stress tensor and fluid temperature that can be applied force from domain along wall boundary $S_w$. The six-node triangular element is used for the development of the finite element equations. All six nodes are associated with velocities, temperature as well as stress tensor; only the corner nodes are associated with pressure. This means that a lower order polynomial is chosen for pressure and which is satisfied through continuity equation. The basic unknowns for the above differential equations are the velocity components $U$, $V$ the stress tensor, $\sigma$ and the pressure, $P$. The velocity component and the stress tensor distributions and linear interpolation for the pressure distribution according to their highest derivative orders in the differential Equations (1-5) as

$$U(X,Y) = N_\alpha U_\alpha \tag{5.10}$$
$$V(X,Y) = N_\alpha V_\alpha \tag{5.11}$$
$$\sigma(X,Y) = N_\alpha \sigma_\alpha \tag{5.12}$$
$$P(X,Y) = H_\lambda P_\lambda \tag{5.13}$$
$$\theta(X,Y) = N_\alpha \theta_\alpha \tag{5.14}$$

Where, $\alpha = 1, 2, \ldots, 6$; $\lambda = 1, 2, 3$; $N_\alpha$ are the element interpolation functions for the velocity components and the stress tensor, and $H_\lambda$ are the element interpolation functions for the pressure. Substituting the element velocity component distributions, the stress tensor distribution, and the pressure distribution from Equations (1-5) the finite element equations can be written in the form,

$$K_{\alpha\beta x} U_\beta + K_{\alpha\beta y} V_\beta = 0 \tag{5.15}$$

$$Re(K_{\alpha\beta\gamma x}U_\beta U_\gamma + K_{\alpha\beta\gamma y}V_\beta U_\gamma) + M_{\alpha\mu x}P\mu + K_{\alpha\beta x}\sigma_\beta + K_{\alpha\beta y}\sigma_\beta$$
$$+ (1-\lambda)(S_{\alpha\beta xx} + S_{\alpha\beta yy})U_\beta - f_x K_\alpha = Q_{\alpha u} \qquad (5.16)$$

$$Re(K_{\alpha\beta\gamma x}U_\beta V_\gamma + K_{\alpha\beta\gamma y}V_\beta V_\gamma) + M_{\alpha\mu y}P_\mu + K_{\alpha\beta x}\sigma_\beta$$
$$+ K_{\alpha\beta y}\sigma_\beta + (1-\lambda)(S_{\alpha\beta xx} + S_{\alpha\beta yy})V_\beta - f_y K_\alpha = Q_{\alpha v} \qquad (5.17)$$

$$W_i(K_{\alpha\beta\gamma x}U_\beta\sigma_\gamma + K_{\alpha\beta\gamma y}V_\beta\sigma_\gamma) + K_{\alpha\beta}\sigma_\mu - \mu_v(K_{\alpha\beta x}U_\alpha + K_{\alpha\beta y}U_\beta)$$
$$-W_i\left(K_{\alpha\beta\gamma x}U_\beta\sigma_\gamma + K_{\alpha\beta\gamma y}U_\beta\sigma_\gamma\right) = Q_{\alpha T} \qquad (5.18)$$

$$Re\,Pr(K_{\alpha\beta\gamma x}U_\beta\theta_\gamma + K_{\alpha\beta\gamma y}V_\beta\theta_\gamma) + (S_{\alpha\beta xx} + S_{\alpha\beta yy})\theta_\beta - K_\alpha Q$$
$$+ \rho_f K_{\alpha\beta}\theta_\beta = Q_{\alpha\theta} \qquad (5.19)$$

Where, the coefficients in element matrices are in the form of the integrals over the element area and along the element edges $S_0$ and $S_w$ as,

$$K_\alpha = \int_A N_\alpha dA \qquad (5.20a)$$
$$K_{\alpha\beta x} = \int_A N_\alpha N_{\beta,x} dA, \qquad (5.20b)$$
$$K_{\alpha\beta y} = \int_A N_\alpha N_{\beta,y} dA, \qquad (5.20c)$$
$$K_{\alpha\beta\gamma x} = \int_A N_\alpha N_\beta N_{\gamma,x} dA, \qquad (5.20d)$$
$$K_{\alpha\beta\gamma y} = \int_A N_\alpha N_\beta N_{\gamma,y} dA, \qquad (5.20e)$$
$$K_{\alpha\beta} = \int_A N_\alpha N_\beta dA, \qquad (5.20f)$$
$$S_{\alpha\beta xx} = \int_A N_{\alpha,x} N_{\beta,x} dA, \qquad (5.20g)$$
$$S_{\alpha\beta yy} = \int_A N_{\alpha,y} N_{\beta,y} dA, \qquad (5.20h)$$
$$M_{\alpha\mu x} = \int_A H_\alpha H_{\mu,x} dA, \qquad (5.20i)$$
$$M_{\alpha\mu y} = \int_A H_\alpha H_{\mu,y} dA, \qquad (5.20j)$$
$$Q_{\alpha u} = \int_{S_0} N_\alpha S_x dS_0, \qquad (5.20k)$$
$$Q_{\alpha v} = \int_{S_0} N_\alpha S_y dS_0, \qquad (5.20l)$$
$$Q_{\alpha\sigma} = \int_{S_0} N_\alpha \sigma_w dS_w \qquad (5.20m)$$
$$Q_\alpha\theta = \int_{S_w} N_\alpha\, q_w dS_w \qquad (5.20n)$$

Using the Newton-Raphson iteration technique the set of nonlinear algebraic Equations (5.15-5.19) are transferred into linear algebraic equations. Finally, these linear equations are solved by applying triangular factorization method and reduced integration technique of Zeinkiewicz and Taylor [40] and the finite element Equations (5.15-5.19) as,

$$F_{\alpha}^{P} = K_{\alpha\beta x} U_{\beta} + K_{\alpha\beta y} V_{\beta} \tag{5.21a}$$

$$F_{\alpha}^{u} = Re(K_{\alpha\beta\gamma x} U_{\beta} U_{\gamma} + K_{\alpha\beta\gamma y} V_{\beta} U_{\gamma}) + M_{\alpha\mu x} P_{\mu} - K_{\alpha\beta x} \sigma_{\beta} - K_{\alpha\beta y} \sigma_{\beta} \\ + (1-\lambda)(S_{\alpha\beta xx} + S_{\alpha\beta yy}) U_{\beta} - f_{x} K_{\alpha} - Q_{\alpha}^{u} \tag{5.21b}$$

$$F_{\alpha}^{v} = Re(K_{\alpha\beta\gamma x} U_{\beta} V_{\gamma} + K_{\alpha\beta\gamma y} V_{\beta} V_{\gamma}) + M_{\alpha\mu y} P_{\mu} - K_{\alpha\beta x} \sigma_{\beta} - K_{\alpha\beta y} \sigma_{\beta} \\ + (1-\lambda)(S_{\alpha\beta xx} + S_{\alpha\beta yy}) V_{\beta} - f_{y} K_{\alpha} - Q_{\alpha}^{v} \tag{5.21c}$$

$$F_{\alpha}^{\sigma} = W_{i}(K_{\alpha\beta\gamma x} U_{\beta} \sigma_{\gamma} + K_{\alpha\beta\gamma y} V_{\beta} \sigma_{\gamma}) + K_{\alpha\beta} \sigma_{\beta} - \mu_{v}(K_{\alpha\beta x} U_{\beta} + K_{\alpha\beta y} V_{\beta}) \\ - 2W_{i}\left( K_{\alpha\beta\gamma x} U_{\gamma} \sigma_{\beta} + K_{\alpha\beta\gamma y} U_{\gamma} \sigma_{\beta} \right) - Q_{\alpha}^{\sigma} \tag{5.21d}$$

$$F_{\alpha}^{\theta} = Re\, Pr(K_{\alpha\beta\gamma x} U_{\beta} \theta_{\gamma} + K_{\alpha\beta\gamma y} V_{\beta} \theta_{\gamma}) + (S_{\alpha\beta xx} + S_{\alpha\beta yy})\theta_{\beta} - K_{\alpha} Q \\ + \rho_{f} K_{\alpha\beta} \theta_{\beta} - Q_{\alpha}^{\theta} \tag{5.21e}$$

This leads to a set of algebraic equations with the incremental unknowns of the element nodal velocity components, temperatures, and pressures in the form,

$$\begin{bmatrix} K_{uu} & K_{uv} & K_{u\sigma} & K_{u\theta} & K_{up} \\ K_{vu} & K_{vv} & K_{v\sigma} & K_{v\theta} & K_{vp} \\ K_{\sigma u} & K_{\sigma v} & K_{\sigma\sigma} & 0 & 0 \\ K_{\theta u} & K_{\theta v} & 0 & K_{\theta\theta} & 0 \\ K_{pu} & K_{pv} & 0 & 0 & 0 \end{bmatrix} \begin{Bmatrix} \Delta u \\ \Delta v \\ \Delta \sigma \\ \Delta \theta \\ \Delta p \end{Bmatrix} = - \begin{Bmatrix} F_{\alpha}^{u} \\ F_{\alpha}^{v} \\ F_{\alpha\sigma} \\ F_{\alpha}^{\theta} \\ F_{\alpha}^{P} \end{Bmatrix} \tag{5.22}$$

Where,

$$K_{uu} = Re(K_{\alpha\beta\gamma x} U_{\gamma} + K_{\alpha\gamma\beta x} U_{\gamma} + K_{\alpha\beta\gamma y} V_{\beta}) + (1-\lambda)\left( S_{\alpha\beta xx} + S_{\alpha\beta yy} \right)$$

$$K_{uv} = K_{\alpha\beta\gamma y} U_{\gamma},\ K_{uT} = -K_{\alpha\beta x} - K_{\alpha\beta y},\ K_{up} = M_{\alpha\mu x},\ K_{vu} = K_{\alpha\beta\gamma x} V_{\gamma},$$

$$K_{vv} = Re(K_{\alpha\beta\gamma x} U_{\beta} + K_{\alpha\gamma\beta x} V_{\gamma} + K_{\alpha\gamma\beta y} V_{\gamma}) + (1-\lambda)\left( S_{\alpha\beta xx} + S_{\alpha\beta yy} \right)$$

$$K_{v\sigma} = -K_{\alpha\beta x} - K_{\alpha\beta y},\ K_{vp} = M_{\alpha\mu y},\ K_{\sigma u} = W_{i} K_{\alpha\beta\gamma x} \sigma_{c} - K_{\alpha\beta x} \mu_{v},$$

$$K_{\sigma v} = W_{i} K_{\alpha\beta\gamma y} \sigma_{c} - K_{\alpha\beta y} \mu_{v},\ K_{\sigma\sigma} = K_{\alpha\beta} - 2W_{i}(K_{\alpha\beta\gamma x} U_{c} + K_{\alpha\beta\gamma y} U_{c})$$

$$K_{\sigma p} = 0,\ K_{pu} = K_{\alpha\beta x},\ K_{pv} = K_{\alpha\beta y}\ \text{and}\ K_{p\sigma} = 0 = K_{pp}.$$

$$K_{\theta\theta} = Re\,Pr(K_{\alpha\beta\gamma^x} U_\beta + K_{\alpha\beta\gamma^y} V_\beta) + (S_{\alpha\beta\,xx} + S_{\alpha\beta\,yy}) + \rho_f K_{\alpha\beta}$$

$$K_{\theta u} = K_{\alpha\beta\gamma^x}\theta\gamma,\ K_{\theta v} = K_{\alpha\beta\gamma^y}\theta\gamma,\ K_{\sigma\theta} = K_{\theta\sigma} = 0$$

$$K_{\theta p} = 0,\ K_{pu} = K_{\alpha\beta^x}\ K_{p\theta} = 0 = K_{pp},\ K_{pv} = K_{\alpha\beta^y}$$

If the percentage of the overall change compared to the previous iteration is less than the specified value, then the iteration process is terminated.

## 4 Numerical experiments

To evaluate the accuracy and confirmation of this work we compute a validation test with Prokop et al [30] for Newtonian (N), generalized Newtonian (GN), Oldroyd-B (OD) and generalized Oldroyd-B (GD) cases. We calculate velocity, pressure and drag coefficient with various dimensionless numbers for considered model. In the present numerical study, the following thermal properties and tissue [29] are considered: $0 \leq Wi \leq 1$, $0 < Re \leq 3000$, $\mu_0 = 0.16$ Pa.s, $\mu_n = 0.0036$ Pa.s, $a = 1.23$, $b = 0.64$, $\lambda = 8.2$s, $\rho = 1050$ kg.m-3, $T_b = 37^0$c, $C_b = 3770$(J/Kg.k), $W_b = 0.5$(Kg/sec.m$^3$), $K = 0.5$(J/s.m.k), $P_f = 400$, $L_w = 2R$, $L = 0.03$m, $R = 3.1$mm, $h_s = R$ and $h_a = 3R$.

It is very important investigation to find out the effects of Reynold numbers, Weissenberg number, wall shear stress, stenotic and aneurysmatic artery, and drag coefficient at stenosed wall on blood flow for the various models Newtonian (N), generalized Newtonian (GN), Oldroyd-B (OD) and generalized Oldroyd-B (GD). The comparative study of blood flow simulation is shown in figures 4, 5 and 7 with the presence and absence of blood clot at the present model in terms of axial velocity and pressure contour lines for all cases. The blood flow characteristics have a significant changed at the throat of stenosis. Figs. 6, 8, 9 and 10 provides the graphical illustration of the velocity, pressure, wall shear stress and drag coefficient of blood flow. The simulation of blood flows has shown in fig. 3 as follows in terms of axial velocity contour lines to code validation [30].

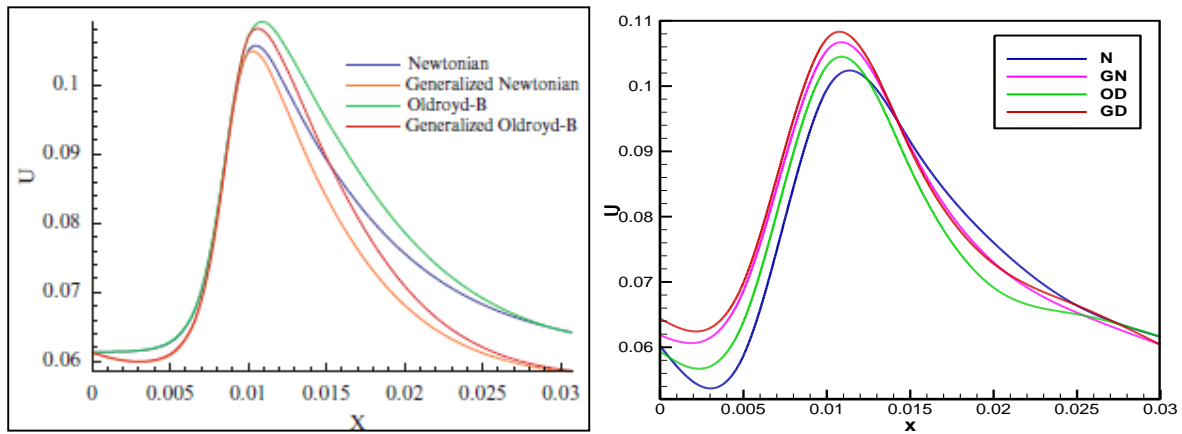

Fig.3 Contrast of velocity contour for all models with V. Prokop et al (left) and present work (right)

## 4.1 Stenotic and Aneurysmatic Effects on blood flow

The blood flow simulation is shown in figs. 4 and 5 with the presence and absence of blood clot at bottom stenosed wall for all models. The indelible recirculation zones are originated at the constriction region of the stenotic and aneurysmatic artery for all cases. The shape of recirculation zone is oval-like and it has reduced at generalized Oldroyd-B model. This recirculation is indicative of iso-blood flow at stenotic artery. The blood shear-thinning properties are important reason to form the recirculation zone and dominate the low-shear area of stenotic and anuerysmatic artery. At aneurysm, the velocity contour lines are almost alike for Newtonian and Oldroyd-B models, but little dissimilarity are found at the rest of models. In the absence of blood clot, the blood flow simulation is presented in fig. 4 (right) for mentioned models where the recirculation zones are found at the throat of stenosis. These recirculation zones are symbolic of regions over a significant portion of each model where the flow is moving with same values. The recirculated area is comparable bigger at blood clotted model than non-blood clotted model for all cases. The shape of recirculation zone is another significant influence of blood clot among the models. At aneurysm, the reversal flow regions and flow separation are found with respect to vessel axis in figs. 4 and 5 but greater flow separation regions are created in non-blood clot model. It is clearly visible for non-blood clot in fig. 5, the back flow is found at just behind the stenotic region and center of aneurysmal area. The more significant disorder of blood flow has developed at dilation area for Newtonian and generalized Newtonian case.

Fig. 6 provides the corresponding effects on blood velocity numerically at the being and absence of blood clot for all models where Re= 1000 and Wi=0.5. It is observed that the velocity profile almost opposite for the blood clot model and non-blood clot model. The maximum blood velocity is found at the throat of stenosis for non-blood clot model and the lowest value is in blood clot model. For the existence of blood clot, the blood velocity is comparable lower to another model which leads local viscosity increases significantly.

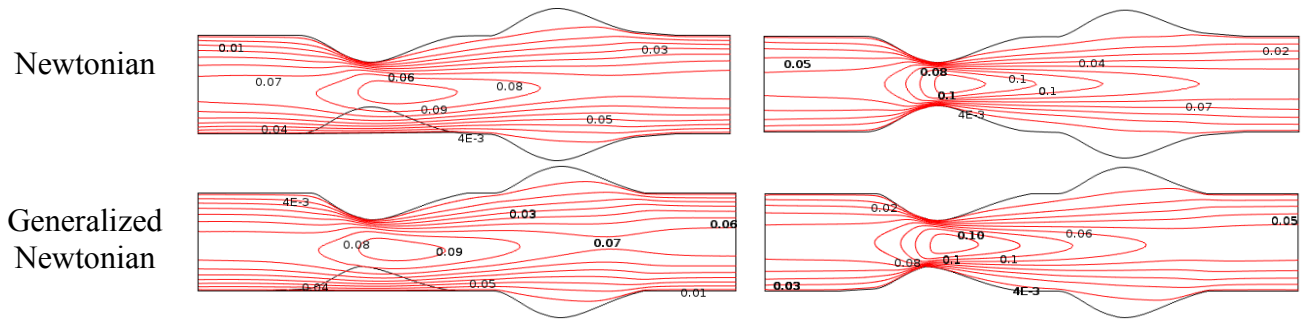

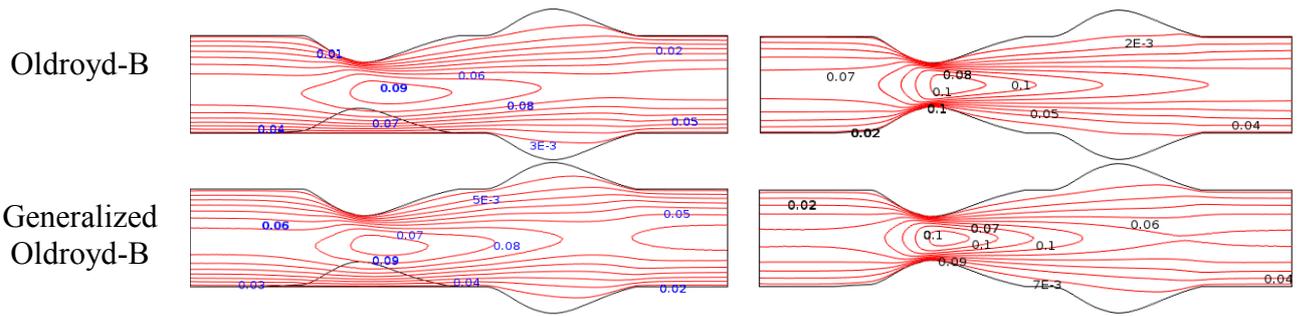

Fig. 4 Velocity contour line on blood flow through stenosed and anuerysmatic vessel with blood clot (left) and without blood clot (right) at *Re*=1000 and *Wi* = 0.6

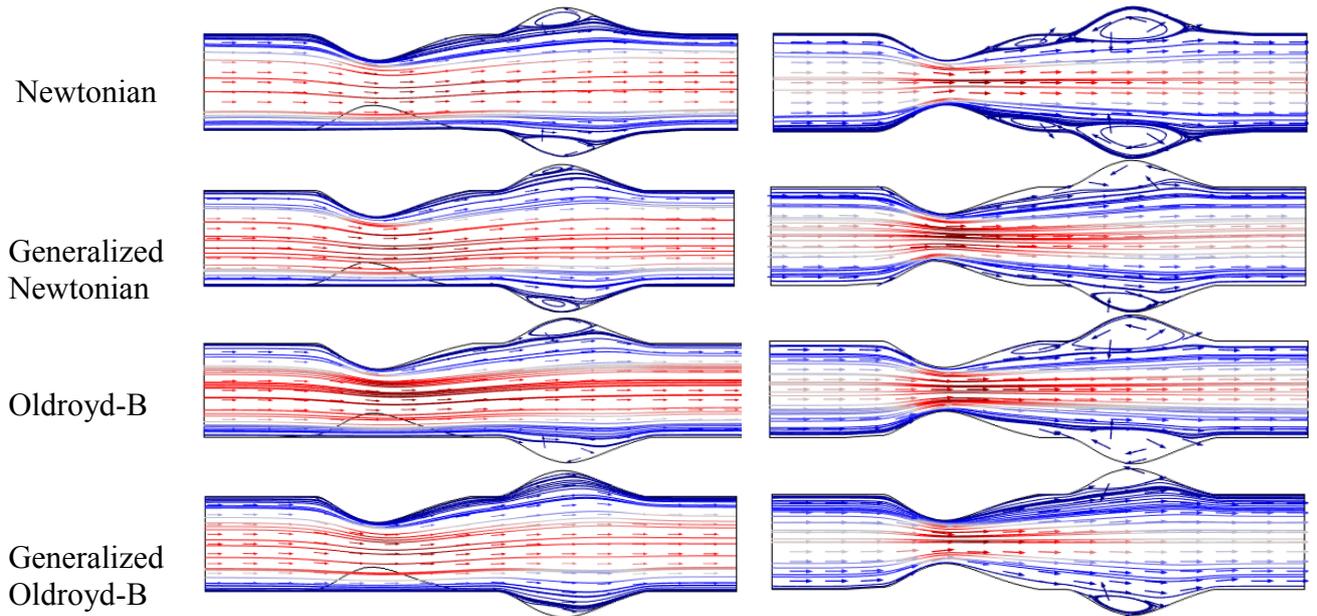

Fig. 5 Vector patterns on blood flow with blood clots (left) without blood clots (right) at *Re*=1000 and *Wi* = 0.6

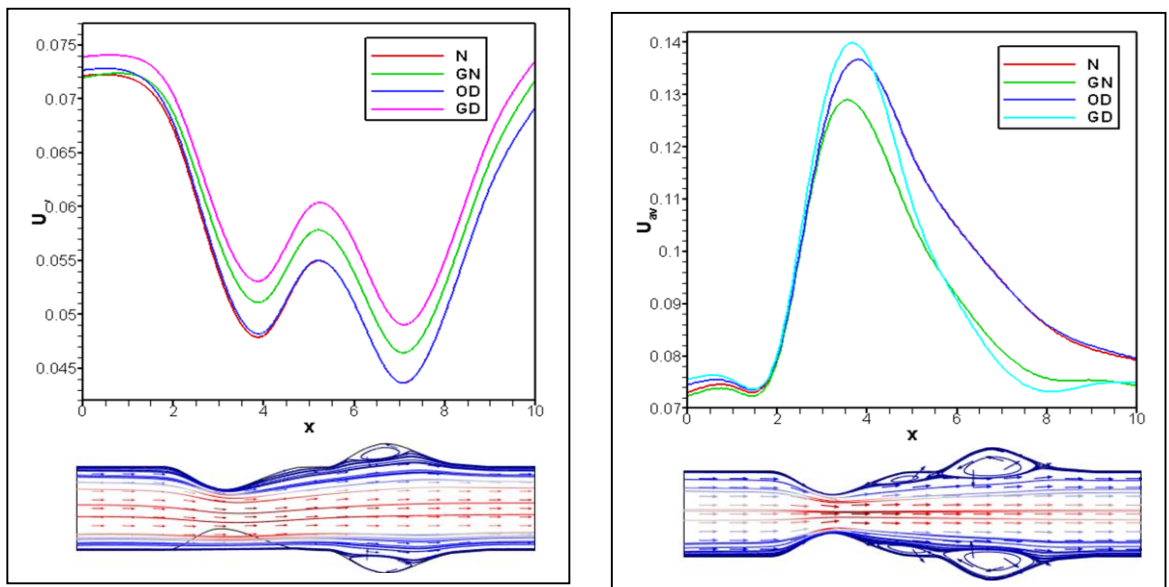

Fig. 6 Comparison of Velocity profile with blood clot (left) and without blood clot (right)

## 4.2 Stenotic and Anuerysmatic effects on pressure distribution

The pressure distribution of blood flow is exhibited for all cases in fig. 7 having blood clot and without blood clot. In fig. 7(left), the steep contour plots display the pressure gained a minimum value at the separation point and decline gradually along vessel axis. Due to formation of blood clots, the pressure peaked at reattachment point at stenosis region and the pressure gradient is high for all events which is agree with Muraki [41] test. The pressure gradient has changed slowly and dense at clotted area. In the case of generalized Oldroyd-B model, the pressure contour lines become more compact and make distort curve within clotted area because of shear-thinning properties of blood. The different pressure contour plots of blood flow are found in fig. 7 (right) for all situations. The parabolic profile has developed at the throat of stenosis and separation points are originated which leads how the pressure reached a minimum point. The main difference is visible between two models at constriction area for all models. The pressure is more dominated at stenosis regions compare to blood clot model because of blood viscosity. In fig. 7 (right), The pressure patterns are almost alike for all models and show the similarities at the far of stenosis, but various pressure contour plots are produces at blood clotted models. From the figures. 8, the blood pressure has decreased gradually in the presence of blood clot and have dramatical changed for non-blood clotting case. In the absence of blood clot, the pressure has gained the lowest value at the throat of stenosis and increased after stenosis. The Newtonian fluid is faster than non-Newtonian fluid for both cases which leads to minimum value at generalized Oldroyd-B model.

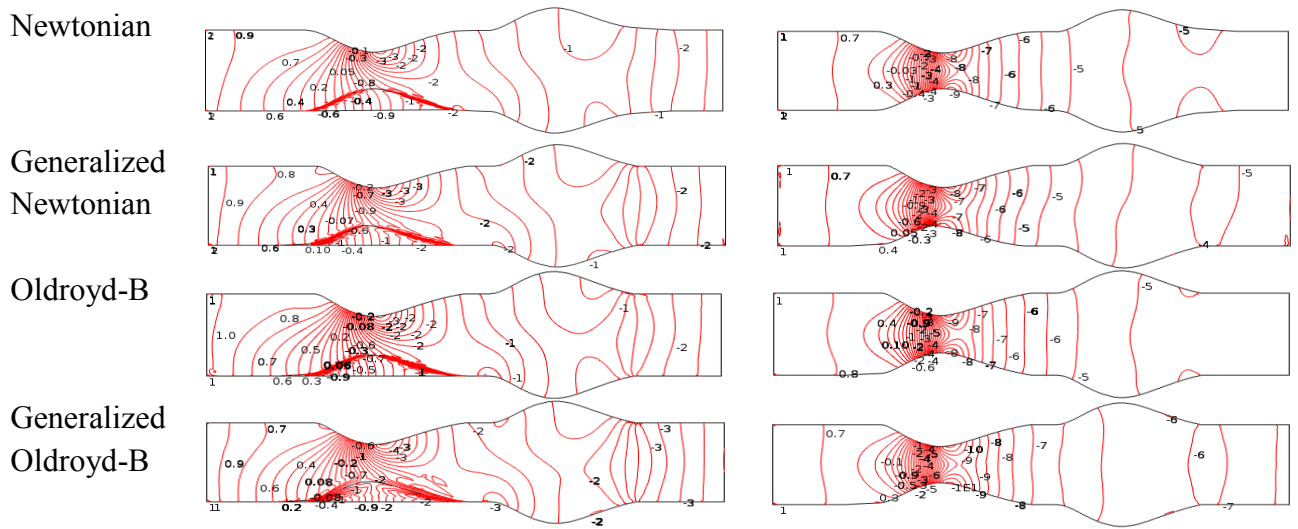

Fig. 7 Pressure distribution on blood flow through stenosed and anuerysmatic vessel with blood clots (left) and without blood clots (right) at $Re=1000$ and $Wi = 0.6$

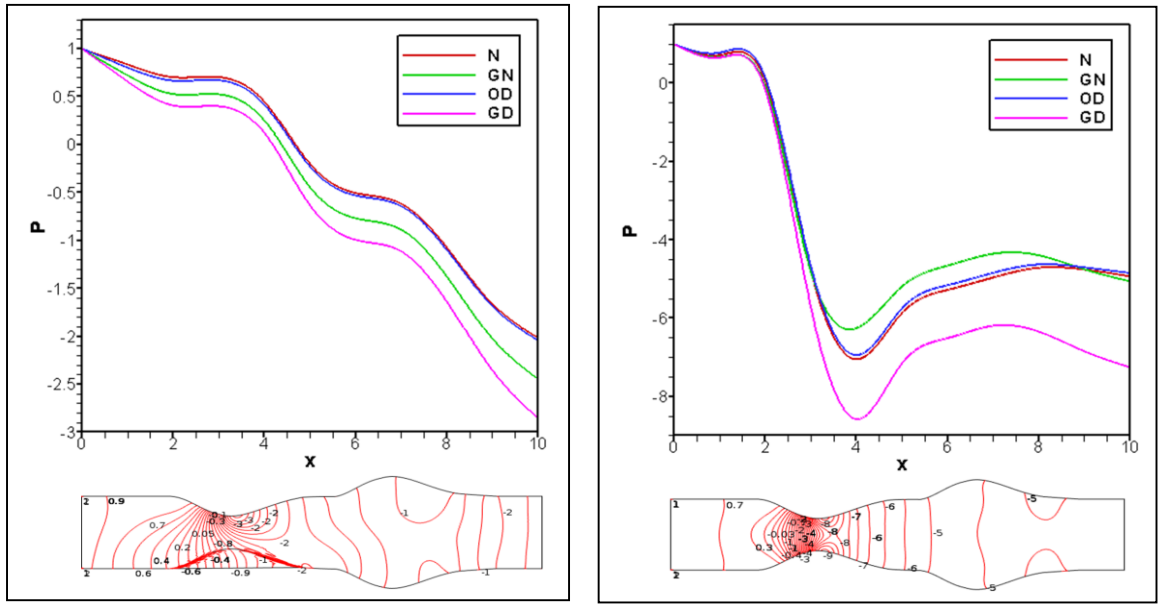

Fig. 8 Comparison of pressure profile with blood clot (left) and without blood clot (right)

### 4.3 Drag coefficient effects on Blood flow

In fig. 9, the drag coefficient (resistance) of blood flow increases at blood clot model and non-blood clot model. It is found that in the absence of blood clots at stenotic area, the drag coefficient is high compare to having blood clots in stenosis artery for Newtonian, Generalized Newtonian, Oldroyd-B, Generalized Oldroyd-B models with the flow rate q=0.1 cm$^3$/s. In the case of non-blood clot, the resistance of blood flow is almost same for Newtonian and Oldroyd-B model but the stenotic zones (by external force) have created more obstacle to blood flow at generalized Oldroyd-B model because of viscoelastic behavior. The hindrance of blood flow over blood lump has increased for all cases but the hurdle is more at generalized model for viscosity. From the fig. 10, the blood flow resistance increases with the increases Reynold numbers and Weissenberg numbers for all four models. The magnitude of drag coefficient is higher for generalized model and lower for Oldroyd-B model with increases of Reynold numbers. Behr et al. [42] have shown that the drag coefficient is a function of Weissenberg numbers and it is increased for all cases with increases of Wi. The most significant value of drag coefficient is originated at generalized Oldroyd-B model in low shear region.

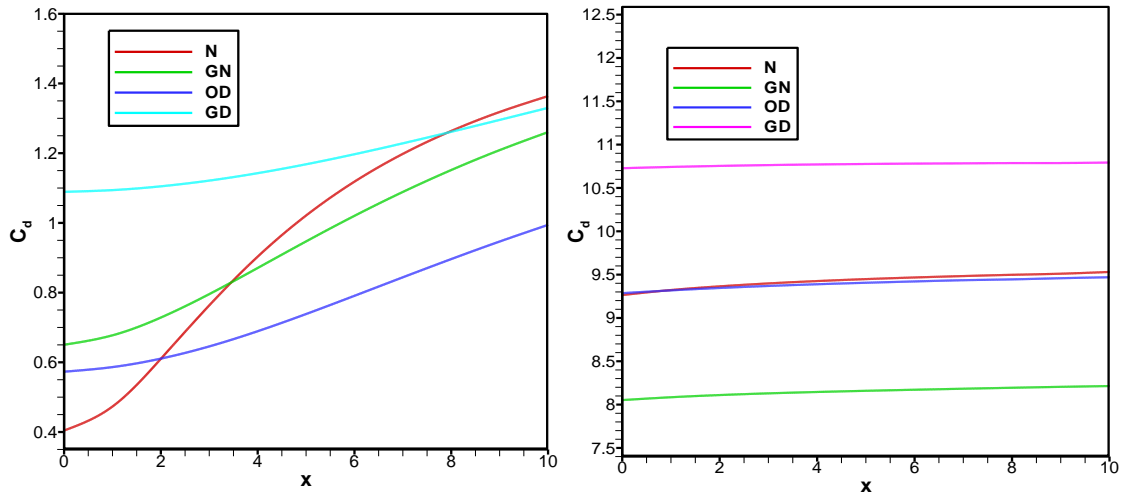

Fig. 9 Drag effects of blood at bottom stenosed vessel wall with blood clots (left) and without blood clots (right) when Re=1000 and Wi=0.6.

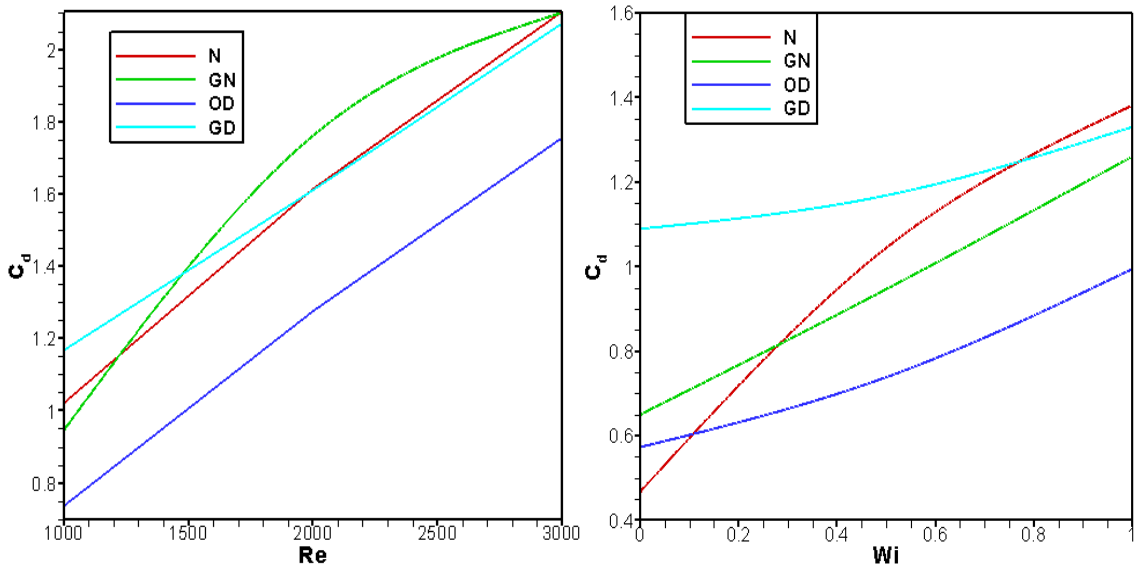

Fig. 10 Drag effects of blood flow at bottom stenosed vessel wall having blood clots with respect to *Re* (left) and *Wi* (right) for all models.

## 5. Conclusion

In this paper we have studied the blood flow patterns through stenotic and anuerysmatic artery having or absence of blood clot for four cases numerically. We have shown that the effect of stenotic vessel, anuerysmatic artery, blood clot, dimensionless number and drag coefficients are very momentous in blood flow. The above factors are correlated to blood viscoelasticity and blood shear thinning behavior. It is found that the blood velocity and pressure have intensive change at the throat of stenosis (no blood clot) compare to blood clot stenosis model.

The blood flow patterns have more affected by high Reynold numbers (turbulence flow) at generalized Oldroyd-B case on the contrary insignificant changed occur for different Weissenberg numbers in present problem. Effects of drag coefficient on blood flow at blood clot regimes are highlighted to explore their impacts on blood flow structure and its characteristics. The efficiency and competence of the theoretical result have been discussed by numerical experiment using finite element technique. Based on the above computational results we conclude that:

a. The shear-thinning effects are related to the blood velocity and pressure and it is more pronounced than the viscoelastic ones.

b. The blood flow parameters are predominant in the recirculation zones compare to blood clot model.

c. The effect of drag coefficient on blood flow is more extreme in stenotic (without blood clot) artery for all models.

d. Due to presence of blood clot at the throat of stenosis the recirculation zones are more elliptic for all models and minimum velocity and maximum blood pressure are observed.

e. The blood velocity has peaked maximum value at narrow cross-sectional area (center of stenotic artery) for absence of blood clot in the model and gained minimum value at wide cross-sectional area with presence of blood clot in the model which is theoretically true. Besides, we have computed blood pressure for present model and the result is quiet agree with theory for all cases.

**Acknowledgements**

We are grateful to the Department of Information and Communication Technology (ICT), Bangladesh University of Professionals (BUP), to provide all facilities during the research.We are grateful to the Department of Information and Communication Technology (ICT), Bangladesh University of Professionals (BUP), to provide all facilities during the research.

**References**

1. Liepsch D.: An introduction to biofluid mechanics basic models and applications. J. Biomech. 35, 415–435 (2002).

2. Nadeem, S., Ijaz, S.: Theoretical analysis of metallic nanoparticles on blood flow through tapered elastic artery with overlapping stenosis. Transaction on Nano biosciences, 14(4), 417-428 (2015).

3. Smith F.T.: The separation flow through a severely constricted symmetric tube. J. Fluid Mech., 90, 725–754 (1979).

4. Belardinelli E., S. Cavalcanti: A new nonlinear two-dimensional model of blood motion in tapered and elastic vessels. Comput. Biol. Med. 21, 1–13 (1991).

5. Long Q., X.Y. Ku, K.V. Ramnarine, P. Hoskins: Numerical investigation of physiologically realistic pulsatile flow through arterial stenosis. J. Biomech. 34 ,1229–1242 (2001).

6. Nerem R.E.: Vascular fluid mechanics, the arterial wall and arteriosclerosis. J. Biomech. Eng. Trans ASME, 114, 274–282 (1992).

7. Cavalcanti S.: Hemodynamics of an artery with mild stenosis. J. Biomech. 28, 387–399 (1995).

8. Chakravarty S., P.K. Mandal: Two-dimensional blood flow through tapered arteries under stenotic conditions, Int. J. Nonlinear. Mech. 35, 779–793 (2000).

9. Mandal P. K: An unsteady analysis of non-Newtonian blood flow through tapered arteries with a stenosis. Int. J. Nonlinear. Mech. 40 151-164 (2005).

10. Mukhopadhyay, S. Layek, G. C.: Analysis of blood flow through a modelled artery with an aneurysm. Applied Mathematics and Computation, 217, 6792-6801(2011).

11. Ingoldby, C. J. H., Wujanto, R., Mitchell, J. E.: Impact of Vascular Surgery on Community Mortality from Ruptured Aortic Aneurysms. Br. J. Surg., 73, 551-563 (1986).

12. Ernst, C.: Abdominal Aortic Aneurysm. N. Engl. J. Med., 328(16), 1167-1172 (1993).

13. Tu C., M. Deville, L. Dheur, L. Vanderschuren: Finite element simulation of pulsatile flow through arterial stenosis. J. Biomech. 25, 1141–1152 (1992).

14. Cronenwett, JL, Murphy, TF, Zelenock, GB, Whitehouse WM, Jr, Lindenauer, SM, Graham, LM, Quint, LE; Silver, TM; Stanley, JC: Actuarial analysis of variables associated with rupture of small abdominal aortic aneurysms, Surgery. 98 (3): 472–83 (1985).

15. Smedby O.: Do plaques grow upstream or downstream? Arterioscler. Thromb. Vasc. Biol. 15, 912–918 (1997).

16. Caro, C. G., Pedley, T. J., Schrote, R. C. R, Seed W. A.: The Mechanics of the Circulation. Oxford University Press, 1978.

17. Porenta G., G.F. Young, T.R. Rogge: A finite element model of blood flow in arteries including taper, branches and obstructions. J. Biomech. Eng. 108, 161–167 (1986).

18. Manton M.J.: Low Reynolds number flow in slowly varying axisymmetric tubes. J. Fluid Mech. 49, 451–459 (1971).

19. Hall P.: Unsteady viscous flow in a pipe of slowly varying cross-section. J. Fluid Mech. 64, 209–226 (1974).

20. Malek J., J. Necas, M. Rokyta, M. Ruzicka: Weak and Measure-Valued Solutions to Evolutionary PDE's. Chapman & Hall, New York, 1996.

21. Rajagopal, K.R., Srinivasa, A.R.: A Gibbs-potential-based formulation for obtaining the response functions for a class of viscoelastic materials. Proc. R. Soc. A, 467, 39-58 (2011).

22. Verdier, C.: Rheological properties of living materials. from cells to tissues. Journal of Theoretical Medicine, 5 (2), 67–91 (2003).

23. Thurston, G.B.: Frequency and shear rate dependence of viscoelasticity of blood. Biorheology, 10(3), 375-381 (1973).

24. Wille, S.: Pulsatile Pressure and Flow in an Arterial Aneurysm Simulated in a mathematical Model. J. Biomed. Eng., 3, 153-158 (1981).

25. Oka, S.: Pressure development in a non-Newtonian flow through a tapered tube. Biorheology, 10(2), 07–212 (1973).

26. Kumar, B. V. R., Naidu, K. B.: Finite element analysis of nonlinear pulsatile suspension flow dynamics in blood vessels with aneurysm. Compur. Biol. Med., 25,1-20(1995).

27. Anand, M. K., Rajagopal, K.R.: A model for the formation and analysis of blood clots. Pathophysiol. Haemost. Thromb., 34, 109–120 (2005).

28. Anand, M., Rajagopal, K. R.: A shear-thinning viscoelastic fluid model for describing the flow of blood. International Journal of Cardiovascular Medicine and Science, 4 (2), 59–68 (2004).

29. Shih, T.C., Yuan P., Lin, W., Kou, H. S.: Analytical analysis of the Pennes bioheat transfer equation with sinusoidal heat flux condition on skin surface. Medical Engineering & Physics, 29, 946– 953 (2007).

30. Prokop, V., Kozel, K: Numerical simulation of Generalized Newtonian and Oldroyd-B Fluids. Numerical Mathematics and Advanced Application 2011, 579-586 (2013).

31. Achab, L., Mahfoud M., Benhadid S.: Numerical study of the non-Newtonian blood flow in a stenosed artery using two rheological models. Thermal Science, 20(2), 449-460 (2016).

32. Uddin N M, Alim M A: Numerical Investigation of Blood Flow through Stenotic Artery. World Journal of Engineering Research and Technology (WJERT), 3(16): 93-116 (2017).

33. Uddin N M, Alim M A: Numerical Study of Blood Flow through Symmetry and Non-Symmetric Stenosis Artery under Various Flow Rates. IOSR Journal of Dental and Medical Sciences, 16(6), 106-115 (2017).

34. Andre P. M.D.: Healthcare extreme, December (2015) https://healthcareextreme.com/how-to-read-your-spine-mri-study/ (accessed March 14, 2020).

35. Ganesh P.: Artery stenosis market outlook, size, growth, share, trends, demand, key players, and regional data statistics, November 6 (2019). https://galusaustralis.com/2019/11/39949/2019-artery-stenosis-market-outlook-size-growth-share-trends-demand-key-players-and-regional-data-statistics/(accessed March 14, 2020).

36. Kumar, B. R., Kumar, G.A., Kumar, S.M.: MATLAB$^R$ and its Application in Engineering. Panjab University, India, 2010.

37. COMSOL Multiphysics, 4.3a users guide, 2013.

38. Dechaumphai, P.: Finite Element Method in Engineering. Bangkok, Chulalongkorn University Press (1999)

39. Taylor, C., Hood, P.: A Numerical Solution of the Navier-Stokes Equations Using Finite Element technique. Computer and Fluids **1**, 73, (1973).

40. Zienkiewicz, O. C., Taylor, R. L.: The finite element method. Fourth Ed., McGraw-Hill. (1991).

41. Muraki, N.: Ultrasonic Studies of the Abdominal Aorta with Special Reference to Hemodynamic Considerations on Thrombus Formation in the Abdominal Aortic Aneurysm. J. Japanese College Angiology, 23,401-413(1983).

42. Behr M., Arora D., Pasquali M.: Stabilized Finite Element Methods of GLS Type for Maxwell-B and Oldroyd-B Viscoelastic Fluids, European Congress on Computational Methods in Applied Sciences and Engineering ECCOMAS 2004, 1-16. (2004)